\documentclass[noinfoline]{imsart}

\RequirePackage[OT1]{fontenc}
\RequirePackage{amsthm,amsmath}
\RequirePackage{natbib}
\RequirePackage[colorlinks,citecolor=blue,urlcolor=blue]{hyperref}

\usepackage{verbatim}
\usepackage{comment}

\allowdisplaybreaks

\usepackage{fullpage}

\newcommand{\E}{\mathbb{E}}
\def\Ez#1{\mathbb{E} \left[ #1 \right]}

\newcommand {\QBartlett}{Q}

\newcommand{\LtwoT}{L^2(\mathcal{T})}
\DeclareMathOperator{\I}{\mathbb{i}}

\DeclareMathOperator*{\argmin}{arg\,min}

\newcommand{\D}[1]{\ensuremath{\operatorname{d}\!{#1}}}

\def\cz#1#2{c^{(#1)}_{#2}}
\def\hcz#1#2{\hat{c}^{(#1)}_{#2}}

\newcommand\Item[1][]{%
  \ifx\relax#1\relax  \item \else \item[#1] \fi
  \abovedisplayskip=0pt\abovedisplayshortskip=0pt~\vspace*{-\baselineskip}}

\usepackage{subfigure}
\usepackage{layout}

\usepackage{dsfont}
\usepackage[mathscr]{eucal}
\usepackage[toc,page]{appendix}
\usepackage{mathrsfs}
\usepackage{color}
\usepackage{pifont}
\usepackage{bm}
\usepackage{latexsym}
\usepackage{amsfonts}
\usepackage{amssymb}
\usepackage{epsfig}
\usepackage{graphicx}
\usepackage{placeins}
\usepackage{multirow}
\usepackage{enumitem}
\usepackage{mathbbol}
\usepackage[dvipsnames]{xcolor}

\newcommand{\transpose}{^{\top}}
\startlocaldefs
\numberwithin{equation}{section}
\theoremstyle{plain}
\endlocaldefs

\begin{document}

\begin{frontmatter}

\title{{\large Yield curve and macroeconomy interaction: evidence from the non-parametric functional lagged regression approach}}

\runtitle{Yield curve and macroeconomy interaction: evidence from the non-parametric functional lagged regression approach}

\begin{aug}
\author{\fnms{Tom{\'a}{\v s}} \snm{Rub{\'i}n}\ead[label=e1]{tomas.rubin@gmail.com}}

\runauthor{T. Rub{\'i}n}

\affiliation{Ecole Polytechnique F\'ed\'erale de Lausanne}

\address{Institut de Math\'ematiques\\
Ecole Polytechnique F\'ed\'erale de Lausanne\\
\printead{e1}}

\end{aug}

\begin{abstract}
Viewing a yield curve as a sparse collection of measurements on a latent continuous random function allows us to model it statistically as a sparsely observed functional time series. Doing so, we use the state-of-the-art methods in non-parametric statistical inference for sparsely observed functional time series to analyse the lagged regression dependence of the US Treasury yield curve on US macroeconomic variables. Our non-parametric analysis confirms previous findings established under parametric assumptions, namely a strong impact of the federal funds rate on the short end of the yield curve and a moderate effect of the annual inflation on the longer end of the yield curve.
\end{abstract}


\end{frontmatter}

\tableofcontents


\section{Introduction}

The yield curve is a collection of yields corresponding to traded debt contracts indexed by varying maturity length (2~months, 1~year, 30~years...), whose construction is well explained by \citet{filipovic2009term}
As an important indicator of the financial sector health, the yield curve is watched closely by traders and investors alike in order to gain understanding about the conditions in financial markets and to discover investment opportunities, and by economists whose analyses provide with conclusions about the economic conditions of the national and global economy, therefore the statistical understanding of the yield curve dynamics is important. The statistical analysis of yield curves is traditionally split into two perspectives: the no-arbitrage approach and the econometric descriptive modelling.

The \textit{no-arbitrage approach} aims to perfectly describe the market data by precisely fitting the term structure in such way that no arbitrage can exist. This fact is quintessential for derivatives pricing formulae that are ultimately based on the same no-arbitrage assumption. Notable contributions in no-arbitrage yield curve modelling include \citet{vasicek1977equilibrium,hull1990pricing,heath1992bond,cox2005theory}.

The \textit{econometric perspective} of yield curve analysis aiming for statistical description of the temporal yield curve evolution has drawn considerably lower attention. \citet{duffee2002term} argued that while the no-arbitrage models admit a good intra-curve fit, they do not depict well the temporal development of the yield curve and such description is necessary for the modelling of the link with the macroeconomy. In their seminal work, \citet{diebold2006forecasting} extended the \textit{Nelson-Siegel factor model} \citep{nelson1987parsimonious} to model the yield curve dynamics. Their state-space framework admits three latent factors, interpreted as level, slope, and curvature, and turned out to be useful for the yield curve forecasting whithin a single market as well as in the interaction analysis among numerous markets \citep{diebold2008global}. Moreover, the framework showed the interaction between the US macroeconomic variables and the US Treasury yield curve \cite{diebold2006macroeconomy}, notably that a positive increase of the federal funds rate, the target rate set by the US Federal Reserve, almost immediately pushes up the slope factor of the yield curve, and a positive increase of the annual inflation influences largely the long-run level of the yield curve.
Similar findings have been also demonstrated by
\citet{rudebusch1999policy,kozicki2001shifting}.

In this article, we take on the econometric approach for the descriptive yield curve modelling and step away from the Nelson-Siegel yield curve parametrisation mentioned in the previous paragraph in favour of the non-parametric domain of \textit{functional data analysis} \citep{ramsay2013functional,ramsay2007applied,hsing2015theoretical}. This statistical discipline considers data sets composed of random functions, or curves, where each of them is treated as an atomic data object and the statistical inference is conducted on an ensemble of such random functions.
The infinite dimensional probabilistic nature of stochastic processes brings over many challenges where
the functional data analysis methodologies need to deviate from the multivariate analysis
methods: the analysis of infinite dimensional problems requires tools from functional analysis
while many standard inference problem may become ill-posed.
The functional data analysis tools have been successfully implemented in modelling handwriting curves \citep{ramsay2000functional}, growth curves \citep{ramsay2013functional}, medical data \citep{ratcliffe2002functionalii,ratcliffe2002functionali,yao2005functional}, and also intra-day trading data analysis \citep{kokoszka2015functional} and volatility modelling \citep{muller2011functional}.

In order to connect the temporal dependence and the non-parametric functional data perspective we consider the \textit{functional time series} perspective conceptualizing the probabilistic model as a temporal sequence of random functions. Many standard univariate and multivariate time series models and concepts have been recently adapted to the functional territory \citep{bosq2012linear,kokoszka2013asymptotic,hormann2010weakly,aue2015prediction,gorecki2018testing,hormann2013functional}.
Moreover, the spectral domain analysis of functional time series data proved to be useful \citep{panaretos2013fourier,panaretos2013cramer,hormann2015dynamic,hormann2015estimation,van2018locally,rubin2020sparsely,rubin2019functional}.
Some classical applications of functional time series analysis include modelling of pollution \cite{hormann2018testing}, DNA dynamics \citep{tavakoli2016detecting}, traffic data \citep{klepsch2017prediction}, or consumer price indexes \citep{chen2016functional}.

Following the parametric model of \citet{diebold2006forecasting} and its aforementioned variants, the yield curves dynamics have been explored by the non-parametric functional time series apparatus in a few following articles. \citet{hays2012functional} estimated the yield curves dynamics by functional dynamic factor framework where the factor loading curves are estimated non-parametrically using smoothness penalisation, while the emphasis of their work was on the factor loading curves interpretation and yield curve forecasting. \citet{kowal2017functional} approached the yield curve modelling implementing functional autoregressive process by the means of Bayesian hierarchical Gaussian models. They derived a Gibbs sampler for inference and forecasting, and conducted an extensive comparative study of yield curve forecasting methods. Finally, \citet{sen2019time} estimated the lag-0 covariance operator of the yield curve by the local-polynomial smoothers and estimated the principal components scores by the \textit{PACE} methodology \citep{yao2005functional} and fitted a vector autoregression.

In this article, we consider the novel spectral domain tools for the functional time series modelling. We consider the US Treasury yield curve as sparsely observed functional time series and estimate the cross-dependence between this data set and the US macroeconomic variables using the local-polynomial smoother techniques \citep{rubin2019functional,rubin2020sparsely}. We model the dependence between the yield curve and the macroeconomic variables by the means of the functional lagged regression \citep{hormann2015estimation,pham2018methodology,rubin2019functional} and estimated the filter-based regression coefficients.
The results of our analysis confirm the findings of \citet{diebold2006forecasting} attained under parametric assumptions. This fact provides with additional supporting argument in favour of the Nelson-Siegel parametric family and might allow for the implementation of the aforementioned parametric models with greater confidence.

The rest of the article is structured as follows. Section~\ref{sec:methodology} establishes the functional data analysis and functional time series methodologies used in our analysis. In particular, Subsection~\ref{subsec:fts} explains the probabilistic analysis of the functional models, Subsection~\ref{subsec:spectral_analysis} summarises the spectral domain results for the functional lagged regression, while Subsection~\ref{subsec:estimation_from_sparse} reviews the estimation methods for sparsely observed functional time series and the estimation of regression filter coefficients. Section~\ref{sec:data_analysis} presents the main findings of this article: the analysis of the US Treasury yield curve dependence on the US macroeconomic variables.
Section~\ref{sec:code} concludes the paper with the link for the code used in this case study.

\section{Methodology}
\label{sec:methodology}

\subsection{Functional time series framework}
\label{subsec:fts}

Functional data analysis \citep{hsing2015theoretical} considers data in the form of random elements in the separable Hilbert space $\mathcal{H}$. The usual choice of such Hilbert space constitutes the Lebesgue space of square integrable functions on the interval $\mathcal{T}$ denoted as $\LtwoT$ with the standard inner product $\langle g_1, g_2 \rangle = \int g_1(x)g_2(x)\D x$ and the norm $\|g_1\| = \langle g_1,g_1\rangle^{1/2}$ for $g_1,g_2\in \LtwoT$. In such case, the sample paths (realisations) of such random elements are assumed to be continuous and smooth.

We shall model the yield curve as a sequence of random functions in $\LtwoT$, referring to such probabilistic object as a \textit{functional time series} and denoting as $\{Y_t(\tau\} = \{ Y_t(\tau), \tau\in \mathcal{T} \}_{t\in\mathbb{Z}}$. We refer .
We additionally assume that the functional time series has finite second moments, $\E \|Y_t\|^2 < \infty$ for all $t\in\mathbb{Z}$, and second-order stationarity in $t$. Therefore we may define the (common) mean function of $\{Y_t(\tau)\}$ by
$$ \mu_Y(\tau) = \Ez{ Y_t(\tau) }, \qquad \tau\in[0,1],$$
and capture the second-order dynamics of the functional time series by its lag-$h$ autocovariance kernels,
$$ R^Y_h(\tau_1,\tau_2) = \E \left\{ (Y_h(\tau_1)-\mu_Y(\tau_2)(Y_0(\tau_2)-\mu_Y(\tau_2)) \right\}, \qquad \tau_1,\tau_2\in[0,1], \quad h\in\mathbb{Z} .$$

The yield curve as a ``platonic'' continuously existing curve is not observed on the full interval: the trading data are available only on a certain finite set of quoted maturities denoted as $\{\tau_1,\dots,\tau_I\} \subset \mathcal{T}$. We consider the data statistically modelled by sampling the functional time series $\{Y_t(\tau)\}$ at the maturities $\tau_1,\dots,\tau_I$ with possible additive noise contamination incorporating small deviations of the quoted data from the smooth model. The considered sampling protocol becomes
\begin{equation}\label{eq:sampling_protocol}
y_{ti} = Y_t(\tau_i) + \epsilon_{ti}, \qquad i=1,\dots,I,\quad t=1,\dots,T,
\end{equation}
where $T$ is the time horizon of the data and $\{\epsilon_{ti}\}$ is an ensemble of  independent identically distributed mean-zero scalar random variables with variance $\sigma^2>0$ which is further assumed to be independent of the functional time series $\{Y_t(\tau)\}$.

Denote the $d$-dimensional multivariate time series of macroeconomic variables as $\{X_t\}_{t\in\mathbb{Z}}$. We shall assume that this time series admits finite second moments and is stationary. Denote its mean vector $\mu_X = \E X_0$ and its elements $\mu_{X^{(j)}} = \E X_0^{(j)}$ for $j=1,\dots,d$, and the lag-$h$ autocovariance matrix
$$\mathbf{R}^X_h = \Ez{ (X_h - \mu_X)(X_0 - \mu_X)\transpose },\qquad h\in\mathbb{Z}.$$
Denoting $( X_t^{(1)},\dots,X_t^{(d)} ) = X_t$ are the elements of the vector of macroeconomic variables,
we propose to model the link between the yield curve functional time series $\{Y_t(\tau)\}$ and the multivariate time series of macroeconomic variables $\{X_\}$ by the lagged regression model \citep{brillinger1981time,hormann2015estimation,rubin2019functional} manifested by the equation
\begin{equation}\label{eq:lagged_regression_model}
Y_t(\tau) = a(\tau) +
\sum_{j=1}^d \sum_{h\in\mathbb{Z}}
b_h^{(j)}(\tau) X_{t-h}^{(j)} + e_t(\tau),\qquad \tau\in\mathcal{T},
\end{equation}
where $a\in\LtwoT$ is called the intercept, $b_h^{(j)} \in \LtwoT$ are the lagged regression coefficients of the macroeconomic variable $j=1,\dots,d$ and lag $h\in\mathbb{Z}$, and $\{e_t\}_{t\in\mathbb{Z}}$ is a sequence of independent identically distributed mean-zero random elements in $\LtwoT$, interpreted as the model error, and which is assumed to be independent of the time series $\{X_t\}$.
For fixed $j$, the coefficients $\{ b_h^{(j)} \}_{h\in\mathbb{Z}}$ are called a filter, and we shall assume their summability in the $\LtwoT$-norm
\begin{equation}\label{eq:summability_filter}
\sum_{h\in\mathbb{Z}} \| b_h^{(j)} \| < \infty, \qquad j=1,\dots,d.
\end{equation}

Finally, define the lagged cross-covariance function between the yield curve $\{Y_t(\tau)\}$ and the macroeconomic variable $\{X_t^{(j)}\}$, for $j=1,\dots,d$, by the formula
$$ R_h^{YX^{(j)}}(\tau) = \Ez{ \left( Y_h(\tau) - \mu_Y(\tau) \right)\left( X_0^{(j)} - \mu_{X^{(j)}} \right) },
\qquad \tau\in\mathcal{T},\quad h\in\mathbb{Z}. $$

\subsection{Spectral analysis of the lagged regression model}
\label{subsec:spectral_analysis}

It turns out that he analysis of the lagged regression model \eqref{eq:lagged_regression_model} becomes simple in the spectral domain.
Under the assumption of the weak dependence, the summability of the lagged autocovariance matrices in the Frobenius norm $\|\cdot\|_F$,
\begin{equation}\label{eq:summability_FX}
\sum_{h\in\mathbb{Z}} \left\| \mathbf{R}^X_h \right\|_F <\infty,
\end{equation}
the spectral density matrix is well-defined \citep{brillinger1981time} by the formula
$$ \mathbf{F}_\omega^X = \frac{1}{2\pi} \sum_{h\in\mathbb{Z}} \mathbf{R}^X_h e^{-\I h\omega}, \qquad \omega\in[-\pi,\pi]. $$
and the lagged autocovariance matrices can be recovered by the inverse formula
$$ \mathbf{R}_h^X = \int_{-\pi}^\pi \mathbf{F}^X_\omega e^{\I h\omega} \D\omega,\qquad h\in\mathbb{Z}. $$
Furthermore, assuming the summubility of the filter coefficients \eqref{eq:summability_filter}, the filter coefficients induce the frequency response function \citep{hormann2015estimation}
\begin{align}
\nonumber
\mathscr{B}_\omega^{(j)}(\tau)  &= \sum_{h\in\mathbb{Z}} b_h^{(j)}(\tau)  e^{-\I k\omega},\qquad\omega\in [-\pi,\pi],\quad j=1,\dots,d, \\
\label{eq:filter_coefficient_inverse_formula}
b_h^{(j)}(\tau)  &= \frac{1}{2\pi} \int_{-\pi}^\pi \mathscr{B}_\omega^{(j)}(\tau)  e^{\I k\omega} \D\omega,\qquad h\in\mathbb{Z},\quad j=1,\dots,d,
\end{align}
Under the model \eqref{eq:lagged_regression_model} and the assumptions \eqref{eq:summability_filter}, \eqref{eq:summability_FX}, the lagged cross-covariance functions satisfies also the weak dependence \citep{rubin2019functional}
$$ \sum_{h\in\mathbb{Z}} \left\| R^{YX^{(j)}}_h \right\| <\infty, \qquad j=1,\dots,d, $$
thus allowing for the definition of the cross-spectral density function
$$ f_\omega^{YX^{(j)}}(\tau) = \sum_{h\in\mathbb{Z}} R_h^{YX^{(j)}}(\tau) e^{-\I h\omega},\qquad \omega\in[-\pi,\pi],\quad \tau\in\mathcal{T},\quad j=1,\dots,d. $$
The lagged regression model\eqref{eq:lagged_regression_model} provides with a simple characterisation in the spectral domain by linking the cross-spectral density function, the frequency response function, and the spectral density matrices
\begin{equation}\label{eq:model_spectral_domain}
f_\omega^{YX}(\tau) = \mathscr{B}_\omega(\tau) \mathbf{F}_\omega^X,\qquad\omega\in[-\pi,\pi],
\end{equation}
where we stuck all components $j=1,\dots,d,$ and define the joint cross-spectral density function $f_\omega^{YX}(\tau) = [ f_\omega^{YX^{(1)}}(\tau) | \cdots | f_\omega^{YX^{(d)}}(\tau) ] $ and the joint frequency response function $\mathscr{B}_\omega(\tau) = [ \mathscr{B}_\omega^{(1)}(\tau) | \cdots |  \mathscr{B}_\omega^{(d)}(\tau) ] $.
The inversion of the spectral domain link \eqref{eq:model_spectral_domain} yields
\begin{equation}\label{eq:model_spectral_domain_solved}
\mathscr{B}_\omega(\tau) = f_\omega^{YX}(\tau) \left( \mathbf{F}_\omega^X \right)^{-1},\qquad\omega\in[-\pi,\pi],
\end{equation}
assuming invertibility of the spectral density matrix $\mathbf{F}_\omega^X$ at each frequency $\omega$, which is a quintessential assumption \citep{brillinger1981time,hormann2015estimation}.

\subsection{Non-parametric estimation of the model components}
\label{subsec:estimation_from_sparse}

The fundamental objective of statistical inference is the estimation of the time-domain and the spectral domain quantities defined in the previous subsection. While the primarily interest lays on the filter coefficients $b_h^{(j)}(\tau)$ identification, as they provide with economic insights into the interaction between the yield curve and the economy, the estimation of the spectral density matrix $\mathbf{F}_\omega^X$ and the cross-spectral density function $f_\omega^{YX}(\cdot)$ constitute mandatory steps towards the frequency response function estimate construction \eqref{eq:model_spectral_domain_solved} and subsequent filter coefficients \eqref{eq:filter_coefficient_inverse_formula}.

The estimation of the marginal properties of the multivariate time series $\{X_t\}_t$ is well established and understood \citep{brillinger1981time}. Considering the sample $X_1,\dots,X_T$, the empirical mean vector is defined by $\hat{\mu}_X = (1/T) \sum_{t=1}^T X_t$ and the empirical lag-$h$ autocovariance matrix by $\hat{\mathbf{R}}_h^X = (1/T) \sum_{t=1}^{T-h} (X_{t+h} - \hat{\mu}_X)(X_{t} - \hat{\mu}_X)\transpose$ for $h=0,1,\dots,T$ and $\hat{\mathbf{R}}_h^X = (\hat{\mathbf{R}}_{-h}^X)\transpose$ for $h=-1,\dots,-T$.

To estimate the spectral density one has to resort to smoothing or a different sort of regularization at some point. \citet{brillinger1981time} suggests kernel smoothing of the periodogram while the alternative \textit{Bartlett's estimate} \citep{bartlett1948smoothing,bartlett1950periodogram} invovles a weighted average of lagged autocovariance matrices with a choice of weights that down-weights higher order lags.
From the theoretical perspective, this approach is equivalent to kernel smoothing of the periodogram \citep[\S 6.2.3]{PriestleyMauriceB1981Saat}. In fact, the Bartlett's weights correspond to the Fourier coefficients of the smoothing kernel, assumed compactly supported.
In this paper, we opt for the Bartlett's estimator with \textit{triangular window} weights defined as $W_h = (1-|h|/\QBartlett)$ for $|h|<\QBartlett$ and $0$ otherwise for the Barlett's span parameter $\QBartlett\in\mathbb{N}$ because of its simplicity. It should be noted that other choices of weights are possible \citep{rice2017plug} and the so-called local quadratic windows (Parzen, Bartlett-Pristley, etc.) improve the asymptotic bias. See \citet[\S 7.5]{PriestleyMauriceB1981Saat} for the detailed discussion in one-dimensional case.
The estimator of the spectral density matrix $\mathbf{F}^X_\omega$ is given by
\begin{equation}
\label{eq:bartlett_matrix}
\hat{\mathbf{F}}_\omega^X = \frac{1}{2\pi} \sum_{h=-\QBartlett}^\QBartlett W_h \hat{\mathbf{R}}_h^X e^{-\I h\omega}, \qquad\omega\in[-\pi,\pi], 
\end{equation}
where the Bartlett's span parameters is usually set at around $\QBartlett \approx \sqrt{T}$.

The challenging part comes from the fact that the yield curve $\{Y_t(\tau)\}$ is observed only indirectly through the data set generated by the protocol \eqref{eq:sampling_protocol}. To overcome this sparse design, we are going to use the local-polynomial smoother based techniques \citep{FanJianqing1996Lpma} that have been used in functional data analysis with great success \citep{yao2005functional,hansen2008uniform,rubin2020sparsely}. While these local-polynomial regression methods are suitable for regression schemes with irregular grids, they can be used also for regular grid \citep{sen2019time} where their purpose lays rather in interpolation. The consistent estimation is indeed unrealistic unless the regular grid grows denser and denser in the limit. Such assumption is not true for the yield curve data where the quoted maturities are more or less fixed. Still, the local-polynomial regression methods benefit from borrowing strength from other data, assuming smoothness, and their interpolation feature is useful for interpretation purposes.

\begin{figure}[b]
\centering
\makebox[\textwidth][c]{
\includegraphics[width=0.99\textwidth]{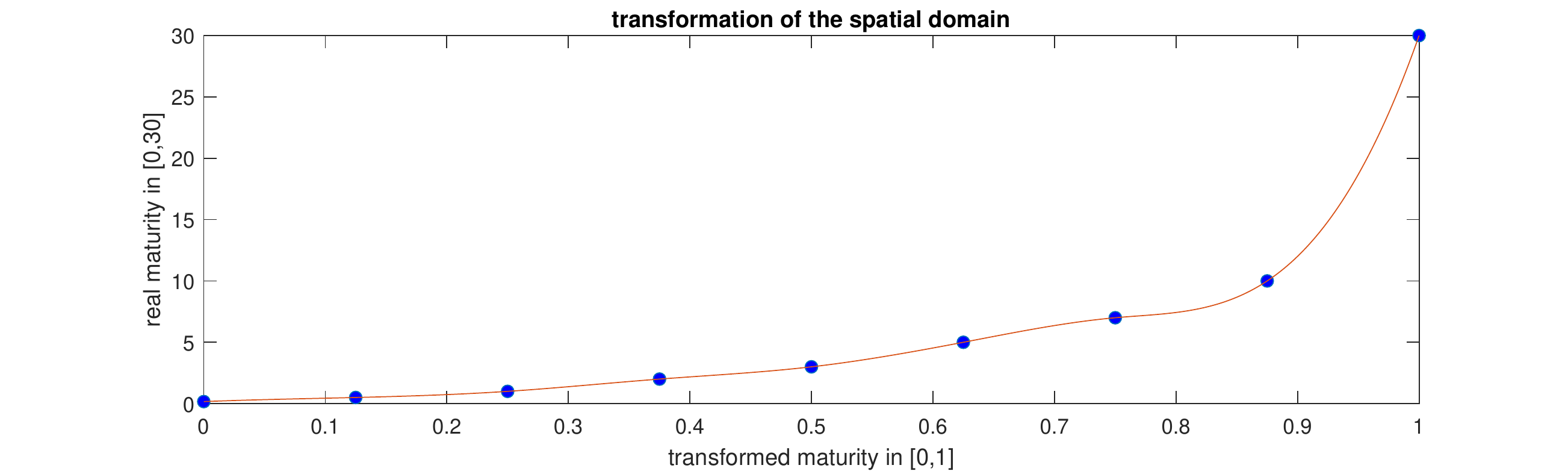}
}
\caption{
Transformation $\varphi : [0,1] \to [0,30]$ arranging the maturities $\tilde{\tau}_i = \varphi^{-1}(\tau_i),\,i=1,\dots,9,$ in order to form an equidistant grid of $[0,1]$.
}
\label{fig5:data3_yield_curve/transformation_maturities}
\end{figure}

The local-polynomial regression methods view the data only locally, where the local neighbourhood is constructed by the weights defined through a fixed kernel function.
Let $K(\cdot)$ be a one-dimensional symmetric probability density function. Throughout this paper we work with the Epanechnikov kernel $K(v)=\frac{3}{4}(1-v^2)$ for $ v\in[-1,1],$ and $0$ otherwise, but any other usual smoothing kernel would be appropriate.
In the usual implementation of the local-polynomial regressions, the data are expected to be scattered more or less uniformly over the spatial domain but this assumption is infeasible as the maturities are rather concentrated close to zero, i.e. the short end of the yield curve. Therefore we define the smooth and strictly increasing bijective transformation $\varphi: [0,1] \to \mathcal{T} $ such that $\tilde{\tau_i} = \varphi^{-1}(\tau_i),\,i=1,\dots,I$, constitute an equidistant partition of $[0,1]$. Such transformation can be constructed, for example, by a monotonous and smooth interpolation of the points $\{( (i-1)/(I-1) , \tau_i)\}_{i=1,\dots,I}$.
Figure~\ref{fig5:data3_yield_curve/transformation_maturities} plots such transformation for the analysed US Treasury yield curve data.

Having established the aforementioned prerequisites, we start with the estimation of the mean yield curve. Define
$\hat{\mu}_Y(\tau) = \hcz{1}{0}(\varphi^{-1}(\tau))$ where, for $\tilde{\tau}\in [0,1]$, $\hcz{1}{0}(\tilde{\tau}) = \hcz{1}{0}$ is obtained by minimizing the weighted sum of squares
\begin{equation}\label{eq:local_LS_for_mu}
\left(\hcz{1}{0},\hcz{1}{1} \right) = \argmin_{\cz{1}{0},\cz{1}{1}} 
\sum_{t=1}^T \sum_{i=1}^{I}
K\left( \frac{\tilde{\tau} - \tilde{\tau}_i}{B_\mu} \right) \left\{ y_{ti} - \cz{1}{0} - \cz{1}{1}(\tilde{\tau} - \tilde{\tau}_i) \right\}^2
\end{equation}
where $B_\mu>0$ is the smoothing bandwidth parameter.

For $j=1,\dots,d$, the cross-spectral density function between is the yield curve $\{Y_t(\tau)\}$ and the $j$-th macroeconomic variable $\{X_t^{(j)}\}$ estimated
\begin{equation}
\label{eq:sparse_univar_specdensity_smoother}
\hat{f}^{YX^{(j)}}_\omega(\tau) = \frac{\QBartlett}{2\pi} \hcz{2}{0}( \varphi^{-1}(\tau) ) \qquad\left( \in\mathbb{C} \right)
\end{equation}
where, for fixed $\tilde{\tau}\in[0,1]$, $\hcz{2}{0}(\tilde{\tau}) = \hcz{2}{0}$ is realised as the minimiser of the following weighted sum of squares
$$
\left(
	\hcz{2}{0},\, \hcz{2}{1}
\right) =
\argmin_{\left( \cz{2}{0},\, \cz{2}{1}\right)\in\mathbb{C}^2}
\sum_{h=-\QBartlett}^{\QBartlett}
\sum_{t=\max(1,1-h)}^{\min(T,T-h)}
\sum_{i=1}^{I}
W_h K\left( \frac{\tilde{\tau} - \tilde{\tau}_i}{B_R} \right)
\left| G^{YX^{(j)}}_{h,t,i} e^{-\I h\omega} - \cz{2}{0} - \cz{2}{1}(\tilde{\tau} - \tilde{\tau}_i)\right|^2
$$
where $B_R>0$ is the smoothing bandwidth parameter and
$$ G^{YX^{(j)}}_{h,t,i} = \left( y_{t+h,i} - \hat{\mu}_Y( \tau_i ) \right) \left( X_t^{(j)} - \hat{\mu}_{X^{(j)}} \right) $$
is called the ``raw'' cross-covariance with $h=-\QBartlett,\dots,\QBartlett,\,t = \max(1,1-h),\dots,\min(T,T-h),\,i=1,\dots,I$.

The solutions to all above least square optimisation problems can be found explicitly by using a standard argument in local-polynomial regression \citep[\S 3.1]{FanJianqing1996Lpma} or \citep[\S B.2]{rubin2020sparsely}. Moreover, the solutions to the spectral density estimator
\eqref{eq:sparse_univar_specdensity_smoother} depend on a handful of terms independent of the frequency $\omega\in[-\pi,\pi]$, that can be precalculated, and multiplication by complex exponentials. This allows a computationally feasible evaluation even on a fine grid of frequencies. 

Putting together $\hat{f}_\omega^{YX}(\tau) = [ \hat{f}_\omega^{YX^{(1)}}(\tau) | \cdots | \hat{f}_\omega^{YX^{(d)}}(\tau) ] $, we have now estimated all components in order to estimate the filter coefficients by the empirical version of the formula \eqref{eq:model_spectral_domain_solved}. It is noteworthy to stress that in contrast to the functional regression models with functional \emph{regressor} time series \citep{hormann2015estimation,pham2018methodology,rubin2019functional}, our setting does not require regularisation of the spectral density matrix inversion $\hat{\mathbf{F}}^X_\omega$, assuming the population spectral density operator $\mathbf{F}^X_\omega$ is invertible for each $\omega\in[-\pi,\pi]$, see \citet{brillinger1981time}. Hence

\begin{equation}\label{eq:frequency_response_estimator}
\hat{\mathscr{B}}_\omega(\tau) = \hat{f}_\omega^{YX}(\tau) \left( \hat{\mathbf{F}}^X_\omega \right)^{-1}, \qquad\omega\in[-\pi,\pi].
\end{equation}
The estimates of the filter coefficients are obtained by integrating the components \eqref{eq:frequency_response_estimator}
\begin{equation}\label{eq:filter_coefficients_estimator}
\hat{b}_h^{(j)}(\tau) = \frac{1}{2\pi} \int_{-\pi}^\pi \hat{\mathscr{B}}_\omega^{(j)}(\tau) e^{\I h \omega} \D\omega,
\qquad h\in\mathbb{Z},\quad j=1,\dots,d,
\end{equation}
where $\hat{\mathscr{B}}_\omega^{(j)}(\tau),\,j=1,\dots,d,$ are the components of $\hat{\mathscr{B}}_\omega(\tau)$.

Having estimated the filter coefficients \eqref{eq:filter_coefficients_estimator} we may perform the prediction of the yield curve given the macroeconomic variables by the formula
$$ \hat{Y}_t(\tau) = \hat{\mu}_Y(\tau) +
\sum_{j=1}^d \sum_{h\in\mathbb{Z}} \hat{b}_h^{(j)}(\tau)
\left( X_{t-h}^{(j)} - \hat{\mu}_{X^{(j)}} \right), \qquad\tau\in\mathcal{T}
$$
where the data outside of our observation window are imputed by the mean value $X_{t} :=  \hat\mu_X$ for $t<1$ or $t>T$. This imputation has however a minimal effect as the filter coefficients $b_h^{(j)}$ vanish as $|h|\to\infty$ usually quickly. In fact, the estimated coefficients $\hat{b}_h^{(j)}$ for the US Treasury yield case study (Section~\ref{sec:data_analysis}) are close to zero already for $h\neq 0$.
In order to access the goodness of fit we suggest to check the $R^2$ coefficient of determination defined by
\begin{align}
\label{eq:R2_coef}
R^2 &= 1 - \frac{SS_{residual}}{SS_{total}}, \\
\nonumber
SS_{residual} &= \sum_{t=1}^T \sum_{i=1}^I \left( y_{ti} - \hat{Y}_t(\tau_i)  \right)^2, \\
\nonumber
SS_{total} &= \sum_{t=1}^T \sum_{i=1}^I \left( y_{ti} - \hat{\mu}_Y(\tau_i)  \right)^2.
\end{align}

\section{Case study: US Treasury yield curve and the US economy interaction}
\label{sec:data_analysis}

The US Treasury yield curve is watched closely by economists and traders alike as it constitutes an important indicator of the US economy's health. Although the linkage of the economy's performance measured through macroeconomic variables has been demonstrated before by statistical methods \citep{rudebusch1999policy,kozicki2001shifting,diebold2006macroeconomy}, the considered analysis consist solely of parametric models. In this section, we regard the yield curves as a sparsely observed functional time series object and apply the nonparametric estimation framework of this thesis. Our objective is to analyse the interaction of the economy and the yield curves without imposing parametric assumptions in order to asses the correctness of the previous results.
We arrive at the same conclusions as the results obtained by assuming the Nelson-Siegel parametric model \citep{nelson1987parsimonious,diebold2006forecasting,diebold2006macroeconomy} and thus confirming the validity and usefulness of the Nelson-Siegel parametrisation when inspecting the link between the economy and the yield curves, and the autoregressive model for the variables and the yield curves evolutions \citep{diebold2006macroeconomy}.

\begin{figure}
\centering
\makebox[\textwidth][c]{
\includegraphics[width=0.99\textwidth]{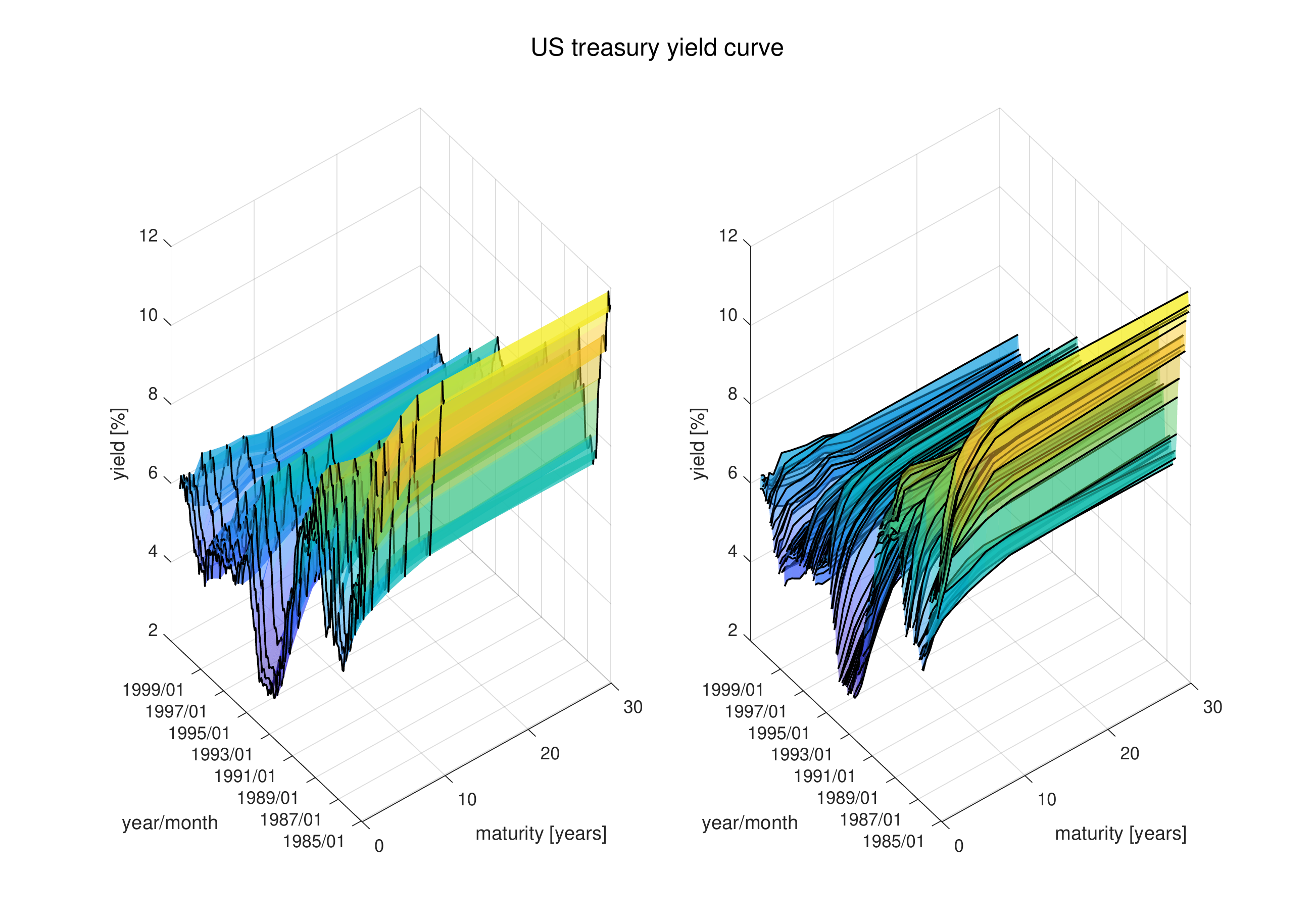}
}
\caption[The US Treasury yield curve]{
The surface on both the left and the right plots displays the evolution of the US Treasury yield curve between the years 1985 and 2000, interpolating the observed yields linearly. \textbf{Left:} the black curves highlight the temporal evolution of the yields at the observed maturities $\tau \in \{ 1/12, 6/12, 1, 2, 3, 5, 7, 10, 30\}$ (years). \textbf{Right:} the black curves constitute the linearly interpolated yields at any given time point, thus plotting the individual yield curves.
}
\label{fig5:data3_yield_curve/yields_surface}
\end{figure}

In order to fully relate the non-parametric viewpoint to the parametric model of \citep{diebold2006macroeconomy} and make the analysis comparable, we consider the same data, namely the U.S. Treasury monthly yield curve between the years 1985 and 2000. In this time interval, the U.S. Treasury issued bonds with the $I=9$ quoted maturities $\tau_i \in \{ 1/12, 6/12, 1, 2, 3, 5, 7, 10, 30\}$ (years). Figure~\ref{fig5:data3_yield_curve/yields_surface} provides with the classical surface plot depicting the yield curves evolution.

\begin{figure}
\centering
\makebox[\textwidth][c]{
\includegraphics[width=0.99\textwidth]{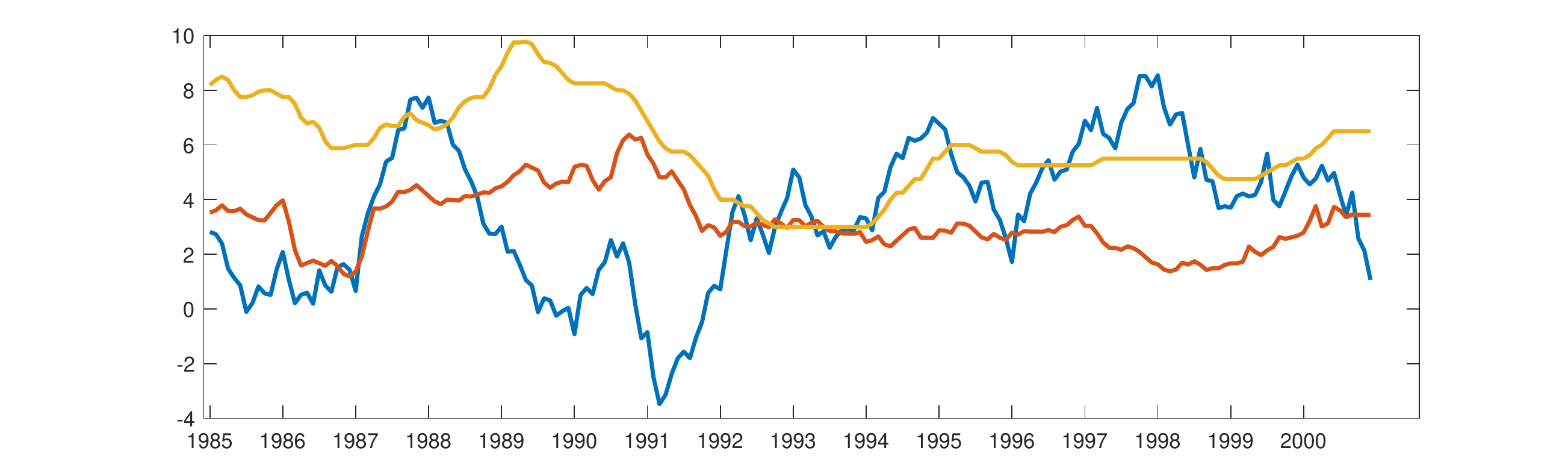}
}
\caption[The macroeconomic variables of the U.S. economy]{
The three macroeconomic time series considered in our analysis.
\textbf{Blue:} Annual change in industrial production [\%].
\textbf{Red:} Annual inflation rate [\%].
\textbf{Yellow:} US federal funds target rate [\%].
}
\label{fig5:data3_yield_curve/macros_plot}
\end{figure}

Moreover, we analysed the following three monthly macro-economic time series (Figure~\ref{fig5:data3_yield_curve/macros_plot}) over the same time period:
\begin{itemize}
\item industrial production index,
\item inflation defined as the annual change of the consumer price index,
\item federal funds rate (interest rate used by banks and other subjects for overnight deposits at the US Federal Reserve).
\end{itemize}

We treat the yield curve as a sparsely observed functional time series. Concretely, $Y_t(\tau), \tau\in[0,30]$ is considered as a random element in $L^2([0,30])$, i.e. setting the interval $\mathcal{T}=[0,30]$.
The interval $[0,30]$ is transformed by a smooth bijective increasing transformation $\varphi$
such that $\tilde{\tau_i} = \varphi^{-1}(\tau_i),\,i=1,\dots,I$, constitute an equidistant partition of $[0,1]$. We construct such transformation $\varphi$ by a cubic spline interpolating the points depicted in Figure~\ref{fig5:data3_yield_curve/transformation_maturities}.
The multivariate time series of macroeconomic variables is denoted as $X_t\in\mathbb{R}^3$, i.e. setting $d=3$.
The time series $\{Y_t(\tau)\}$ and $\{X_t\}$ are assumed to satisfy the assumptions listed in Section~\ref{sec:methodology}, in particular the lagged regression model \eqref{eq:lagged_regression_model}.

\begin{figure}
\centering
\makebox[\textwidth][c]{
\includegraphics[width=0.99\textwidth]{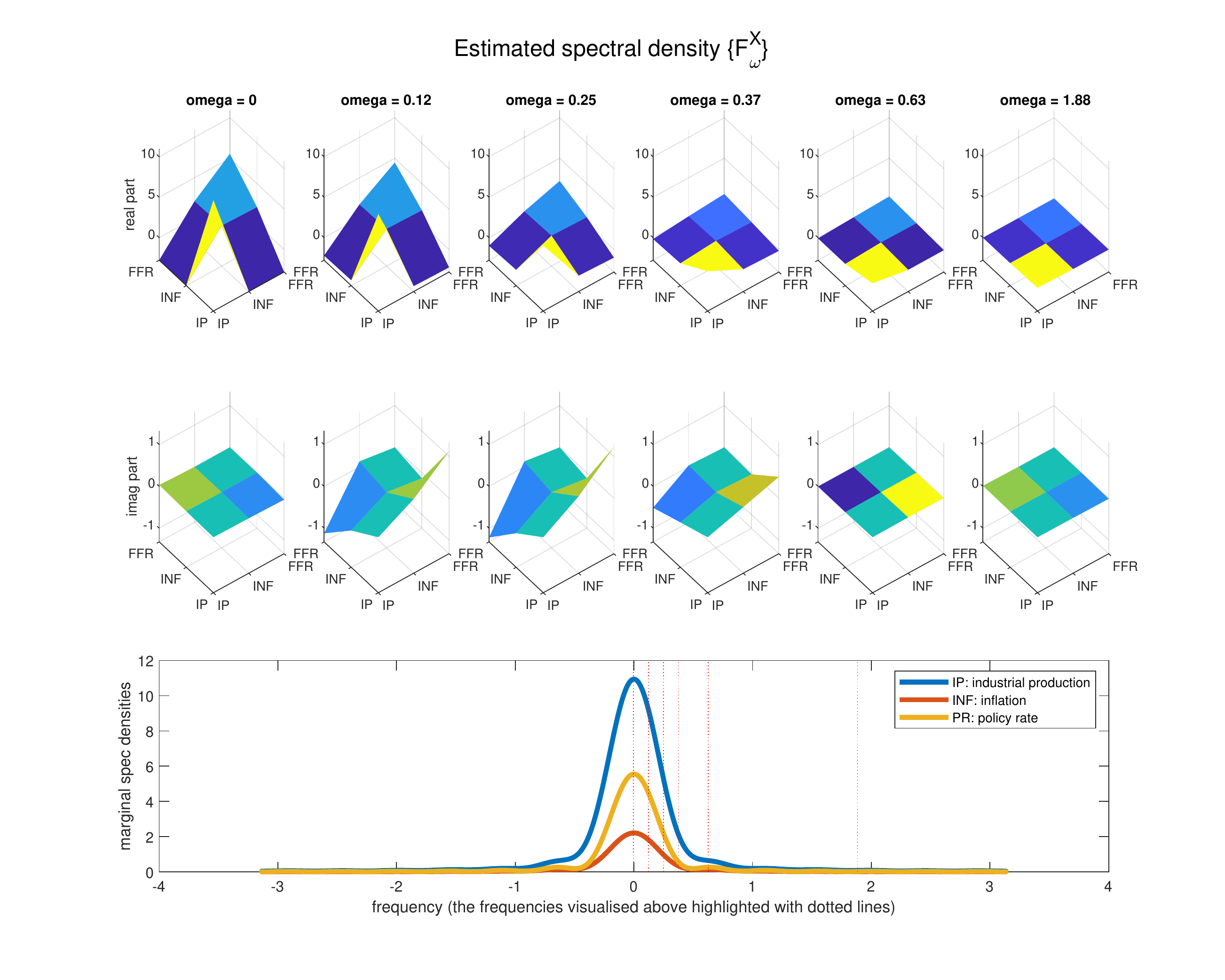}
}
\caption[The spectral density of the US macroeconomic variables]{
A visualisation of the estimated spectral density matrices $\{\hat{\mathbf{F}}_\omega^X\}$ corresponding to the macroeconomic variable time series $\{X_t\}$. \textbf{Top and center rows:} The real and the imaginary parts of the empirical spectral density matrix $\hat{\mathbf{F}}_\omega^X$ evaluated at six different frequencies.
\textbf{Bottom:} The marginal spectral densities of the individual macroeconomic variables which coincide with the diagonal elements of the spectral density matrices $\{\hat{\mathbf{F}}_\omega^X\}$. The dotted vertical lines denote the frequencies visualised above.
}
\label{fig5:data3_yield_curve/macros_spec_density}
\end{figure}

Starting with the spectral analysis of the macroeconomic variables $\{X_t\}$ time series, we estimate the spectral density matrices by the Bartlett's formula \eqref{eq:bartlett_matrix} with Bartlett's span parameter $L = \lceil \sqrt{192} \rceil = 14$. Figure~\ref{fig5:data3_yield_curve/macros_spec_density} visualises the estimated spectral density matrices. The estimated shape of the spectral densities is typical for autoregressive processes with the autoregressive parameter being close to one. This observation might be the first indication that the parametric model of \citet{diebold2006macroeconomy}, in particular the autoregressive model for the temporal evolution, seems to be appropriate.

\begin{figure}
\centering
\makebox[\textwidth][c]{
\includegraphics[width=0.99\textwidth]{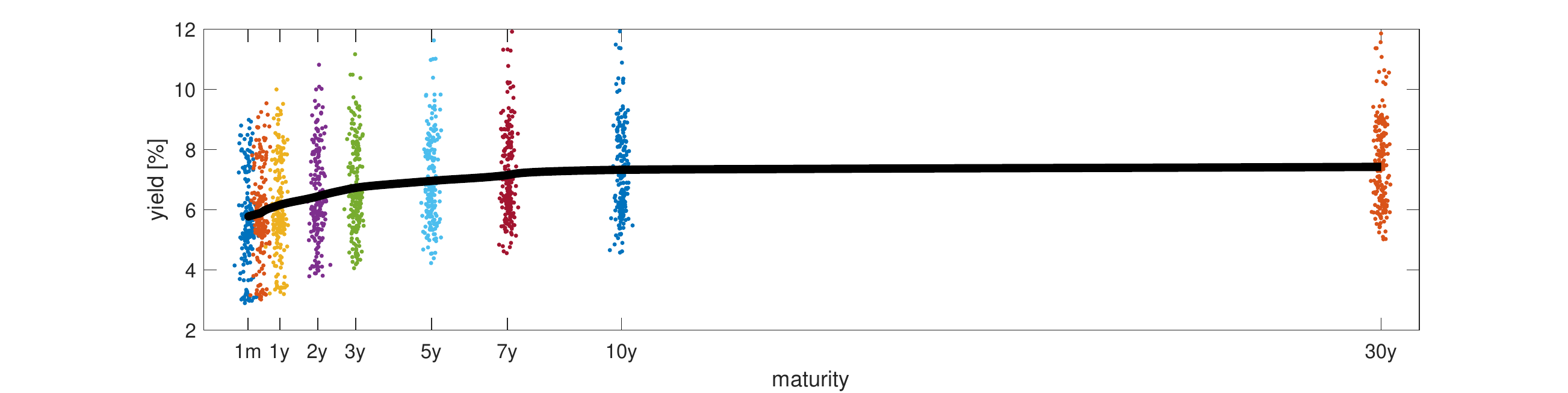}
}
\caption[The estimated mean yield curve]{
The estimated mean yield curve $\{Y_t(\tau)\}$ (the black line) obtained by the local-linear line smoother. The data clouds of individual colours constitute the jittered yields at given maturity.
}
\label{fig5:data3_yield_curve/yields_mean_curve}
\end{figure}

Secondly, the mean yield curve $\mu_Y$ is estimated by the local-linear smoother \eqref{eq:local_LS_for_mu}. The estimate is visualised on Figure~\ref{fig5:data3_yield_curve/yields_mean_curve}.

\begin{figure}
\centering
\makebox[\textwidth][c]{
\includegraphics[width=0.99\textwidth]{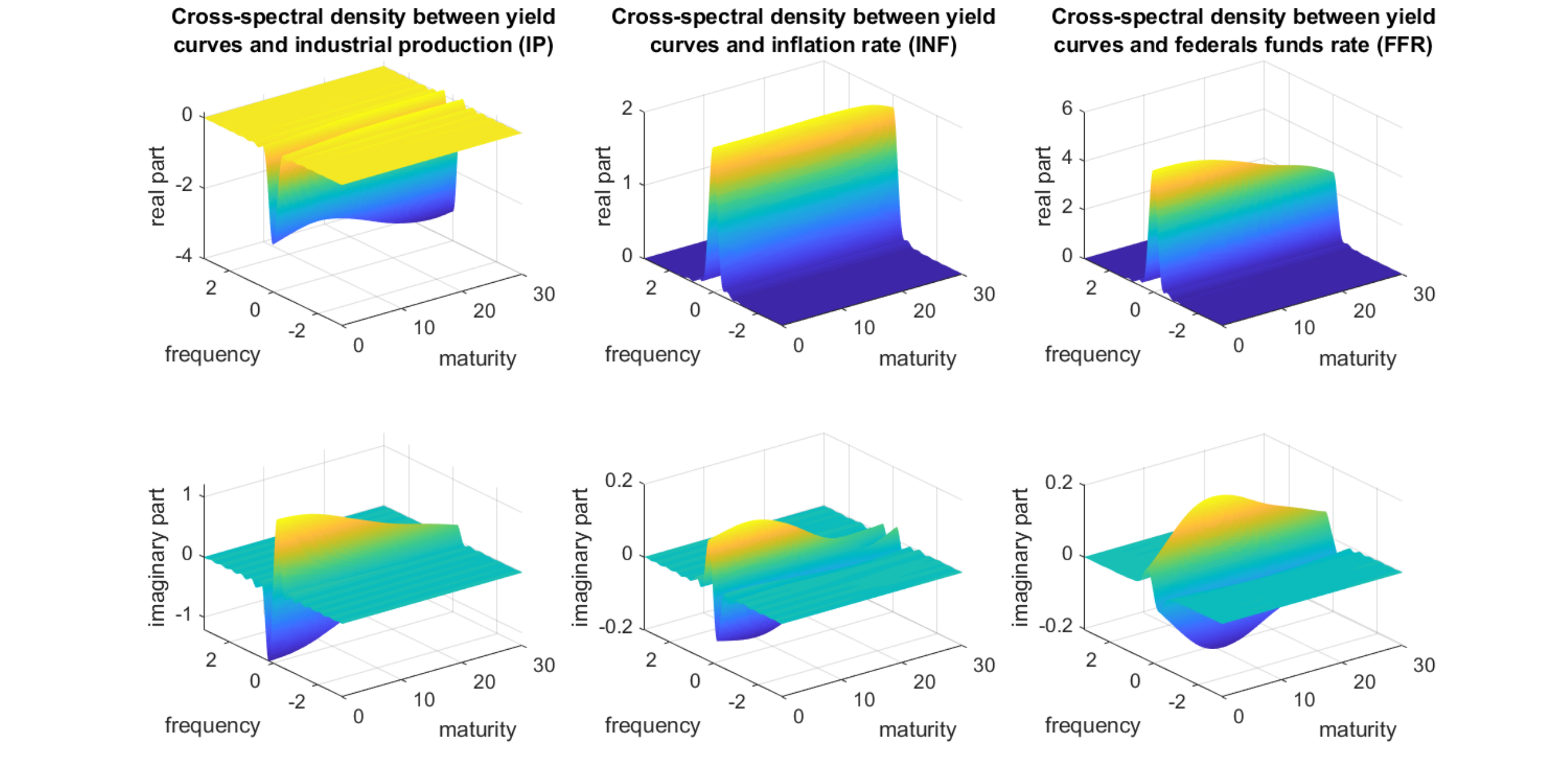}
}
\caption[The estimated cross-spectral density between the US Treasury yield curves and the macroeconomic variables]{
The estimated cross-spectral functions $(\omega,\tau) \mapsto \hat{f}_\omega^{YX^{(j)}}$ between the yield curves functional time series $\{Y_t(\tau)\}$ and the economic variables time series $\{X^{(j)}_t\}$, for the industrial production (IP, $j=1$), inflation rate (INF, $j=2$), and federal funds rate (FFR, $j=3$).
}
\label{fig5:data3_yield_curve/cross_spectral_density}
\end{figure}

Thirdly, we estimate the cross-spectral density functions $f_\omega^{YX}(\tau)$ between the functional time series $\{Y_t(\tau)\}$ and the vector time series $\{X\}$ by the local-liner smoother \eqref{eq:sparse_univar_specdensity_smoother}, again with the Bartlett's span parameter is again set by $L=\lceil \sqrt{192} \rceil = 14$.
Figure~\ref{fig5:data3_yield_curve/cross_spectral_density} visualises the estimated cross-spectral density functions $\hat{f}_\omega^{YX}(\tau)$.
The cross-spectral density functions provides with no informative interpretation, it is rather a tool to obtain the filter coefficients estimates.

\begin{figure}
\centering
\makebox[\textwidth][c]{
\includegraphics[width=0.99\textwidth]{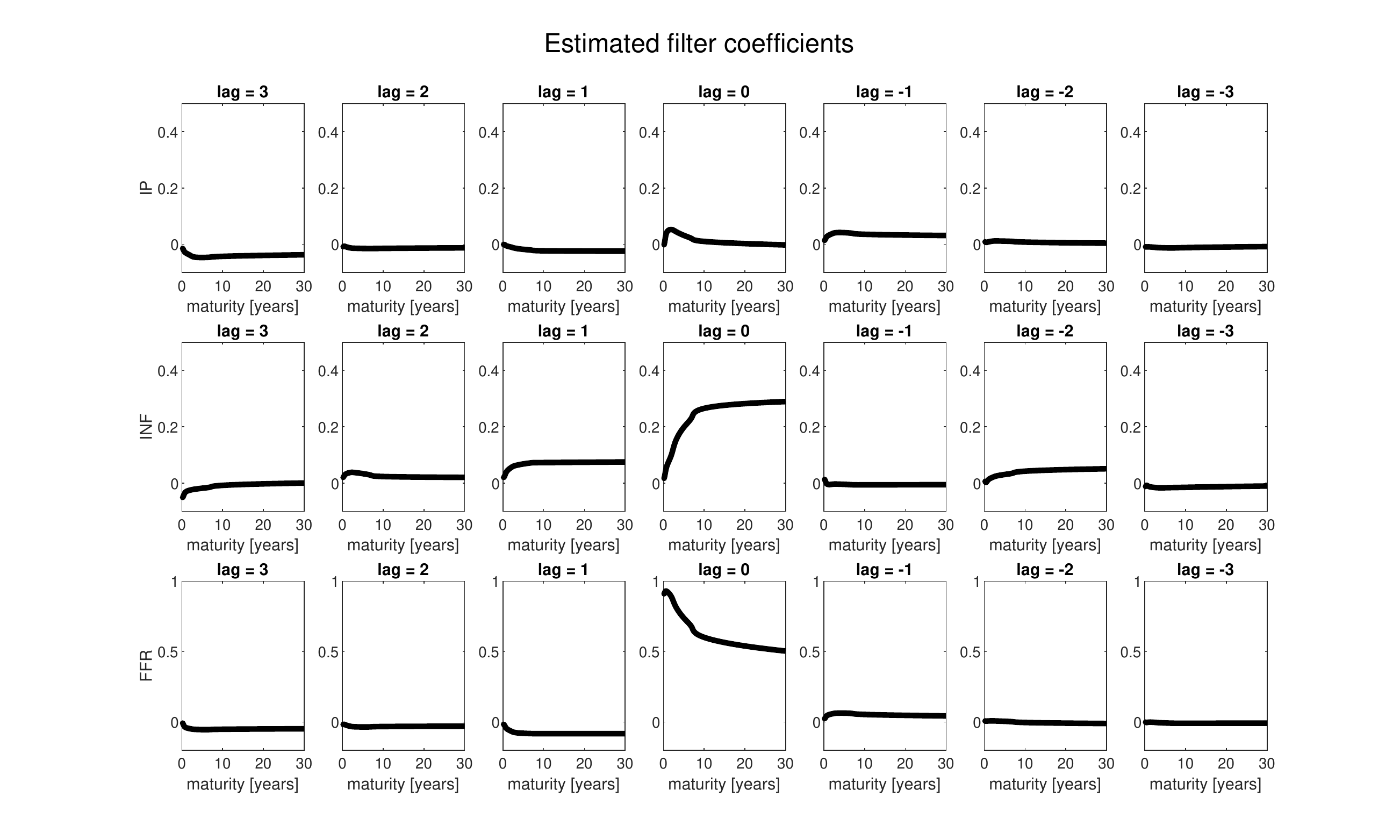}
}
\caption{
The estimated filter coefficients $\hat{b}_h^{(j)}$ for $h=-3,-2,\dots,3$ and the macroeconomic variables: the industrial production (IP, $j=1$), inflation rate (INF, $j=2$), and federal funds rate (FFR, $j=3$).
}
\label{fig5:data3_yield_curve/filter_coefficients}
\end{figure}

Finally, the frequency response function $\mathscr{B}_\omega$ is estimated by the formula \eqref{eq:frequency_response_estimator} and the filter coefficient estimators are obtained by the integration into the temporal domain \eqref{eq:filter_coefficients_estimator}. 
Figure~\ref{fig5:data3_yield_curve/filter_coefficients} shows the estimates of the filter coefficients $\hat{\mathscr{B}}_k$ for $k=-3,-2,\dots,3$. The interpretation of the estimated filter coefficients is the following:
\begin{itemize}
\item The changes of the macroeconomic variables have only imminent impact on the yield curve, the impact is not delayed.
\item The industrial production index (IP) has minimal impact on the yield curves, nevertheless a positive increase of the industrial production seems to increase the yield curve at the shortest of maturities by a tiny bit.
\item The annual inflation (INF) seems to influence the yield curve more. Concretely, an increase of the inflation results into the increase of the yield curve at higher maturities but hardly has an effects on the short maturities. 
\item The federal funds rate (FFR) is linked with the yield curve the strongest among the considered macroeconomic variables. The linkage is the most profound at the short maturities, in particular the filter coefficient function reaches the value close to one at short maturities, signifying that the short end of the market driven yield curve follows closely the federal funds rate. The impact of the federal funds rate is nevertheless significant also for the longer maturities.
\end{itemize}
Our conclusions reflect the same findings as \citet{diebold2006macroeconomy} who studied the macroeconomic interactions with the yield curve using a parametric model, modelling the yield curve using the Nelson-Siegel parametric family \citep{nelson1987parsimonious,diebold2006forecasting} and assuming a vector autoregressive model of the temporal yield curve evolution on the factors level. 

The $R^2$ coefficient of determination \eqref{eq:R2_coef} of such model evaluation gives a quite high value of 0.78, proving a good fit of the model \eqref{eq:lagged_regression_model}.

\FloatBarrier

\section{Code availability}
\label{sec:code}

The code in Matlab that supports the finding of this case study is openly available as a GitHub repository via the link
\url{https://github.com/tomasrubin/us-yield-curve-macroeconomics}. 

\section*{Acknowledgements}
I thank Victor M. Panaretos for guidance on submitting this article.

\bibliographystyle{plainnat} 
\bibliography{bibliography}

\end{document}